# Tunnel magnetoresistance due to Coulomb blockade effects in quasi-one dimensional polymer nanofibers

H. J. Lee[1], A. N. Aleshin[1,2,*], S. H. Jhang[1], H. S. Kim[1], M. J. Goh[3], K. Akagi[3], J. S. Brooks[4] and Y. W. Park[1]

[1]*Department of Physics and Astronomy & Nano Systems Institute - National Core Research Center, Seoul National University, Seoul 151-747, Korea*
[2]*Ioffe Physical-Technical Institute, Russian Academy of Sciences, St. Petersburg 194021, Russia*
[3]*Departmentof Polymer Chemistry, Graduate School of Engineering, Kyoto University, Kyoto 615-8510, Japan*
[4]*National High Magnetic Field Laboratory, Florida State University, Tallahassee, Florida 32306, USA*

We report on the low temperature tunnel magnetoresistance (MR) in quasi one-dimensional (1D) nanofibers made of conjugated polymers. The MR voltage bias dependence reveals an enhancement (at low biases) and the oscillatory behavior at temperatures below 10 K. The low temperature isotropic MR behavior has been attributed to the charging effects in the polymer nanofiber which considered as an array of small conducting regions separated by nano barriers. These effects at low temperatures lead to the single electron tunneling represented by the Coulomb blockade regime as well as to an enhancement and oscillation of the tunnel MR.

Interplay of electrical charging and spin-dependent tunneling effects in small metallic granular systems gives rise to significant magnetotransport phenomena in such nanostructures. The charging effect of small metallic islands results in Coulomb blockade regime below a threshold voltage, $V_t$, due to single electron tunneling (SET) [1]. SET phenomena are characteristic feature of inorganic nanostructures consist of conducting islands separated by tunnel barriers, which results in multiple junctions along an array. It is worth noting that in such structures spin-dependent SET affects strongly the low temperature tunnel magnetoresistance (MR) by leading to an enhancement and oscillation of the MR. This phenomenon has been predicted theoretically [2,3] and proved experimentally, in particular, for ferromagnetic granular systems [4,5]. Obviously, the restriction of the current path is an important condition in order to observe the SET phenomena. This condition can be accomplished in such inherent quasi one-dimensional (1D) systems as conducting nanofibers made of conjugated polymers. Such nanofibers demonstrate the transport features characteristic of quasi-1D systems at temperature 30 K < T < 300 K [6-8], whereas at T < 30 K the observed current-voltage (I-V) behavior and the differential conductance fluctuations have been attributed to Coulomb blockade effects [9]. By analogy to inorganic nanowires [10] such a doped polymer fiber can be considered at low temperature as an array of small conducting islands separated by nanoscale barriers (Fig. 1) where the Coulomb blockade tunneling is the dominant transport mechanism [9]. It is evident that in the Coulomb blockade regime the application of strong magnetic field should affect such a low dimensional nanoarray of conducting islands by leading to the tunnel MR (TMR). In this case one might expect an oscillation of such a TMR similar to that found in double ferromagnetic tunnel junctions [2,3].

In this study we report on the low temperature MR in quasi-1D polymer nanofibers. The voltage bias dependence of the MR reveals an enhancement and the oscillatory behavior at temperature below 10 K. The low temperature MR behavior has been attributed to the charging effects in the polymer nanofiber which considered as an array of small conducting regions separated by nanoscale barriers. These effects lead to a single electron tunneling represented by the Coulomb blockade regime as well as by an enhancement and oscillation of the TMR.

We have studied helical PA (hel PA) nanofibers synthesized using chiral nematic liquid crystal as a solvent of Ziegler-Natta catalyst following the procedure presented elsewhere [11]. Helical PA can be made with either R- (counterclockwise) or S- (clockwise) type helicity. Single R-hel PA fibers with a cross-section, typically 40 - 60 nm (high), 100 - 300 nm (wide) and the length up to 10 μm, were deposited on a Si substrate with a $SiO_2$ layer and platinum electrodes thermally evaporated on the top, 2 μm apart. The fibers were doped with iodine from vapor phase up to the saturation level similar to our previous studies [6-9]. The transport measurements were performed in the 2-probe geometry in a vacuum at ~$10^{-5}$ Torr, using the cryostat with a superconducting magnet and a Keithley 6517A electrometer. Magnetic field up to 18 T was applied either parallel or perpendicular to the sample current direction using the rotator at the sample holder. MR was evaluated from the difference between the I-V curves measured at H = 0 T and 18 T as MR = [R(H) – R(0)]/R(0).

Figure 1 shows an Atomic Force Microscope (AFM) image of typical R-hel PA nanofiber used in our study on top of Pt electrodes. Figure 2 presents the I-V behavior for

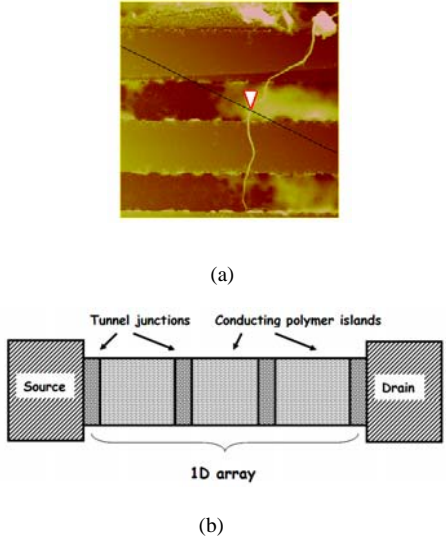

(a)

(b)

Fig. 1. (a) AFM image of typical R-hel PA nanofiber used in our study I-V characteristics on top of Pt electrodes; (b) Schematic of the metal-insulator-metal tunneling junction structure of a conducting polymer nanofiber.

such R-hel PA nanofiber at T = 1.4 K at magnetic fields, H = 0 and 18 T, perpendicular to the current direction. As can be seen from Fig. 2, the low temperature I-Vs are symmetric and strongly non-linear in both cases: with and without magnetic field. Inset 1 to Fig. 2 demonstrates that there is the threshold voltage, $V_t$, below that the current is zero, in particular, for this fiber the $V_t$ ~ 0.45 V at T = 1.4 K. At $V < V_t$ the Coulomb blockade regime occurs similar to that found in our previous study [8,9], while the current increases rapidly when voltage bias exceeds $V_t$. For $V > V_t$ current scales as $(V/V_t - 1)^\zeta$, with $\zeta$ ~ 1.8 – 1.9 at intermediate biases, characteristic of transport in 2D nanoarrays of conducting islands separated by nanobarriers [9]. The remarkable Coulomb blockade effect may be associated with the fact that the tunneling paths are limited in such nanoarrays as polymer nanofibers and the charging energy of conducting islands is large enough. Inset 2 to Fig. 2 shows the typical behavior of the magnetoconductance (MC) plotted as function of magnetic field for the R-hel PA fiber under discussion. We found that the MC is negative (MR is positive) and highly noisy. Figure 3a shows the bias voltage dependence of the MR estimated from the I-V curves at T = 1.4 K for a R-hel PA fiber shown in Fig. 2. As can be seen from Fig. 3a, positive MR increases with decreasing bias voltage and reaches the maximum value at the voltage bias slightly above $V_t$. At $V < V_t$ the MR is not shown because of absence of current in the Coulomb blockade regime. Surprisingly we have observed the MR oscillations as a function of bias voltage. This behavior is characteristic of the tunnel MR and it contradicts with the theory prediction for nonferromagnetic systems such as polymer fiber deposited onto Pt electrodes, where the MR should be flat

[2]. An estimated period of the observed MR oscillations at T = 1.4 K is

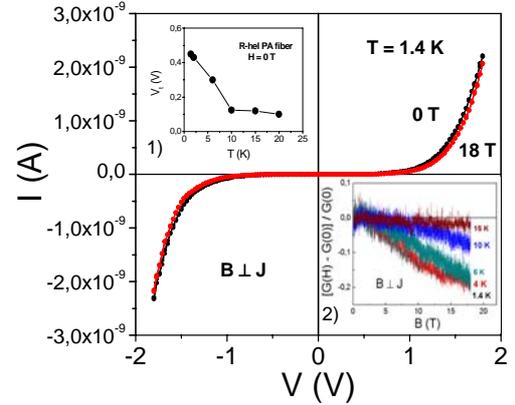

Fig. 2. I-V characteristics for R-hel PA nanofiber at T = 1.4 K at H = 0 and 18 T with a magnetic field perpendicular to the current direction. Inset 1: temperature dependence of the threshold voltage, $V_t$, for the same fiber. Inset 2: typical magnetoconductance vs B of the R-hel PA nanofiber as B perpendicular to the current direction..

approximately ~ 200 – 250 mV (see Fig. 3a) – of the same order of magnitude as the period of differential conductance oscillations for R-hel PA fibers found in our previous study [9]. The period of the MR peaks supposed to repeat with the period of expected Coulomb staircases despite these staircases are not well pronounced in the I-Vs for the same sample (Fig. 2). Figure 3b shows that the magnitude of the transverse MR and their oscillations decreases rapidly with increasing temperature and almost disappears at T = 10 K, where the charging energy of the conducting islands becomes comparable with $k_B T$. As can be seen from Figure 4, the anisotropy of MR in magnetic field parallel and perpendicular to the current is rather small probably because of complicated helical structure, which makes the tunneling probability in different directions more or less equal. At the same time the MR oscillations are slightly higher for the transverse MR configuration. Qualitatively similar results were obtained for several other R-hel PA fibers.

It is evident that SET phenomena manifest themselves if the charging energy $E_C = e^2/2C$ is much larger than the thermal energy $k_B T$, where C is the junction capacitance [1]. This means that the junction area or the size of isolated islands should be small enough to observe SET phenomena in polymer nanofibers. One can suggest that the applied magnetic field affects the I-V curve that causes an enhancement of the MR at the unobservable (due to measuring system threshold) steps of the Coulomb staircase resulting in the oscillation of the MR. The observed MR behavior for R-hel PA nanofibers is qualitatively consistent with the theoretical predictions for the TMR enhancement and its oscillation in double tunnel junctions caused by the effect of spin-dependent SET [2,3]. The magnitude of the TMR oscillation observed in R-hel PA fiber is relatively small since the height of the tunnel barriers between conducting polymer islands in the polymer nanoarray is rather low. The presence of several tunneling junction nanoarrays in parallel in the polymer nanofiber and the randomness of junction capacitances may also smear the TMR oscillation especially at high temperature. It is worth noting that the similar behavior including the TMR oscillations in electric fields has been theoretically predicted for double (or multiple) ferromagnetic junctions [2,3]. Such a structure should consist of a metallic island separated from both electrodes by insulating barriers and at least one of the electrodes should be ferromagnetic. The discrete charging of the metallic islands leads to negative MR as well as to oscillations of the TMR with increasing voltage. As theory predicts in the case of modest charging energy the bias dependence of the TMR includes two components: a term which varies smoothly with increasing bias voltage and a term with an oscillatory behavior [2]. The oscillatory component vanishes when the thermal energy becomes of the order of the charging energy or when spin asymmetries of both junctions are the same. This effect has been found and studied thoroughly in Co-Al-O thin films, which are ferromagnetic indeed [3,4]. We realize that negative TMR is not common for ferromagnetic junctions with identical contacts, however such MR behavior has been predicted and explained theoretically for asymmetric, inhomogeneous junctions and seems to be an indication of the coherent tunneling [12,13].

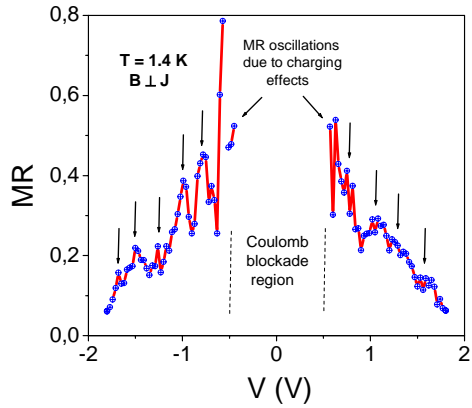

(a)

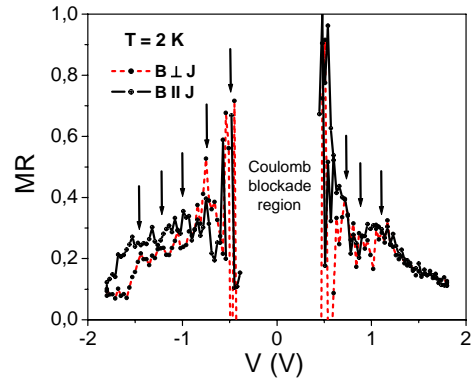

Fig. 4. Bias voltage dependence of MR at T = 2 K for the same R-hel PA fiber at magnetic field perpendicular and parallel to the current direction.

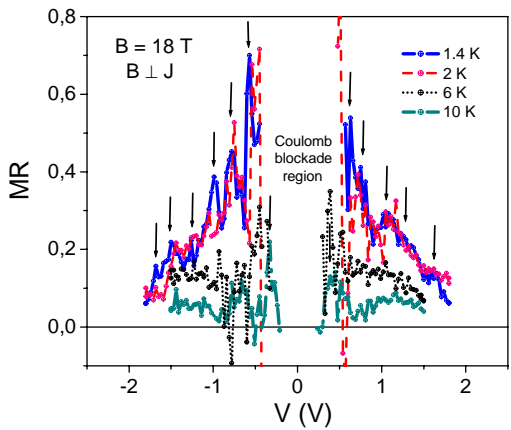

(b)

Fig. 3. (a) (a) Bias voltage dependence of MR at T = 1.4 K for the same R-hel PA fiber; (b) Bias voltage dependence of MR for the same R-hel PA fiber at different temperatures, T, K: 1.4; 2; 6; 10.

Qualitatively the MR behavior observed in R-hel PA nanofibers coincides with theoretical models [12,13]. Rather noisy MC behavior (Inset 2 to Fig. 2) may indicate the multiple hopping or tunneling processes between conducting islands. The oscillatory behavior of the MR is present but not well pronounced because of very complex internal structure of helical PA fibers. There is the only difference between our results and theories for the TMR. Namely, the materials used in our study, R-hel PA nanofibers deposited between Pt electrodes, are paramagnetics [14]. The origin of the TMR oscillations and thus spin-dependent SET in the R-hel PA nanofiber, which consists of low dimensional nanoarrays of conducting islands between Pt electrodes, is not completely clear at this moment. It was shown that the modest doping of the PA chain results in the formation of charged or neutral solitons. The former carry charge without spin, the latter does not carry charge but has spin which can provide the spin-dependent effects [15]. Iodine doping affects transport in the PA chain by creation of conjugation defects, domain walls (soliton kinks), etc. This results in the appearance of conductive islands with presence of neutral solitons (with spin). The size of such islands should be comparable to the distance between impurities ~ 10 nm [9]. The array of such islands may act as multiple asymmetric, inhomogeneous tunnel junctions. On the other hand the magnetic domains responsible for the TMR oscillations might originate from some impurities came from the catalyst residue. It is worth noting that Ziegler-Natta catalyst [$AlR_3$/$Ti(OR)_4$] used for synthesis of R-hel PA contains such nonferromagnetic atoms as Al and Ti. [14]. The TMR-like behavior in nonferromagnetic tunnel junctions has been found recently in Ti/$TiO_2$/Ti structures [16], which is similar to that observed in R-hel PA nanofibers. Therefore, the presence of residue atoms of Ti in conducting PA nanofibers may affect the MR leading to its enhancement and oscillation in the Coulomb blockade regime similar to that in TMR structures. Moreover the period of dI/dV oscillations in Ti/TiO$_2$/Ti tunnel junctions is ~ 250 – 300 mV that is very close to that for the MR oscillations in R-hel PA fibers. Thus one may expect that the MR oscillations at low temperature could arise from spontaneous formation of Ti/TiO$_2$ domains with "frozen" spin orientation inside the polymer nanoarray, which can act as multiple asymmetric, inhomogeneous tunnel junctions.

In conclusion we have investigated the low temperature MR in quasi-1D polymer nanofibers. The voltage bias dependence of the MR shows an enhancement (at low biases) and the oscillatory behavior which becomes more pronounced as temperature decreases below 10 K. We attributed the low temperature isotropic MR behavior to the charging effects in the polymer nanofiber which considered as an array of small conducting regions separated by nanoscale barriers. The charging effects lead to a single electron tunneling represented by the Coulomb blockade regime observed in quasi-1D polymer nanofibers as well as to an enhancement and oscillation of the TMR at low temperature. The observed behavior characteristic of spin-dependent transport may originate from both: lightly doped islands of PA nanofibers with presence of neutral solitons and residue impurities like Ti arise from the Ziegler-Natta catalyst, which both can act as asymmetric, inhomogeneous tunnel junctions. We believe that further experimental and theoretical studies are necessary to clarify the origin of the tunnel MR behavior observed in quasi-1D polymer fibers.

We thank V.I. Kozub, M. Fogler and S.N. Artemenko for valuable comments. This work was supported by the Nano Systems Institute – National Core Research Center (NSI-NCRC) program of KOSEF, Korea. Support from the subprogram of the Presidium of Russian Academy of sciences "Fundamental principles of the developing and investigation of new substances and materials for molecular electronics and spintronics" for A.N.A. is gratefully acknowledged. A portion of this work was performed at the National High Magnetic Field Laboratory, supported by NSF Cooperative Agreement No.DMR-95-27035 and by the State of Florida, USA.